\documentclass[%
 reprint,
 amsmath,amssymb,
 aps,
floatfix,
]{revtex4-2}

\usepackage{hyperref}% add hypertext capabilities
\newcommand{\fullcirc}{\mbox{{\Large$\bullet$}}}

\newcommand{\fullsquare}{\mbox{$\blacksquare$}}

\newcommand{\fulltriangle}{\mbox{$\blacktriangle$}}

\usepackage{graphicx}% Include figure files
\usepackage{dcolumn}% Align table columns on decimal point
\usepackage{bm}% bold math
\usepackage{color}

\hypersetup{
	breaklinks=true,   % splits links across lines
}

\begin{document}

\preprint{APS/123-QED}

\title{Breather Solutions to a Two-dimensional \\ Nonlinear Schr{\"o}dinger Equation with Non-local Derivatives}% Force line breaks with \\

\author{Alexander Hrabski }
\author{Yulin Pan}%
 \email{yulinpan@umich.edu}
\affiliation{%
Department of Naval Architecture and Marine Engineering, University of Michigan, Ann Arbor, Michigan 48109, USA
}%

\date{\today}

\begin{abstract} %<=600 characters. Currently 599.
We consider the nonlinear Schr{\"o}dinger equation with non-local derivatives in a two-dimensional periodic domain. For certain orders of derivatives, we find a new type of breather solution dominating the field evolution at low nonlinearity levels. With the increase of nonlinearity, the breathers break down, giving way to wave turbulence (or Rayleigh-Jeans) spectra. Phase-space trajectories associated with the breather solutions are found to be close to that of the linear system, revealing a connection between the breather solution and Kolmogorov-Arnold-Moser (KAM) theory.

%\begin{description}
%\item[Usage]
%Secondary publications and information retrieval purposes.
%\item[Structure]
%You may use the \texttt{description} environment to structure your abstract;
%use the optional argument of the \verb+\item+ command to give the category of each item. 
%\end{description}
\end{abstract}

%\keywords{Suggested keywords}%Use showkeys class option if keyword
                              %display desired
\maketitle

%\tableofcontents

\section{Introduction}

Breathers are a broadly-defined class of features that arise in nonlinear dynamical systems, generally describing a family of solutions with strong spatial localization and oscillations in time. Together with solitons, breathers are considered as prototypes for rogue waves that can occur across many fields, such as water waves \cite{dysthe_note_1999,onorato_rogue_2013}, optics \cite{dudley_instabilities_2014}, and plasma physics \cite{ding_breathers_2019}. Mathematically, breathers are fundamental solutions to both continuous field equations and discrete lattice problems. In the former case, breather solutions have been found primarily in one-dimensional (1D) nonlinear partial differential equations, including the Sine-Gordon equation \cite{cuevas-maraver_sine-gordon_2014}, the nonlinear Schr{\"o}dinger equation (NLS) \cite{tajiri_breather_1998,dysthe_note_1999,onorato_rogue_2013,dudley_instabilities_2014,ding_breathers_2019}, and the Korteweg–De Vries (KdV) equation \cite{clarke_generation_2000,chow_interactions_2005}. In the latter case, discrete breathers \cite{flach_discrete_2008} (as counterparts to breathers in continuous fields) have been constructed as solutions to a wide variety of systems, including Josephson Junctions \cite{trias_discrete_2000,miroshnichenko_breathers_2001} and the Fermi-Pasta-Ulam-Tsingou (FPUT) problem \cite{livi_breathers_1997}. Unlike their continuous counterparts, discrete breathers haven been studied extensively in problems with more than one dimension.

Another important category of studies regards the spontaneous emergence of breathers (and other types of coherent structures) under the free evolution of a system. These coherent structures include the quasi-solitons \cite{zakharov_wave_2001,rumpf_turbulent_2009}, quasi-breathers \cite{pushkarev_quasibreathers_2013}, and wave collapses \cite{zakharov_wave_2001,rumpf_wave_2013,rumpf_transition_2015} identified in the 1D Majda-McLaughlin-Tabak (MMT) model, as well as the discrete breathers in the FPUT problem \cite{cretegny_localization_1998} and the discrete nonlinear Schr{\"o}dinger equation \cite{rumpf_coherent_2001,rumpf_simple_2004}. As in the case of constructing exact breather solutions, when considering continuous fields, these studies are predominantly performed for 1D situations. The only exception, to our knowledge, is \cite{saint-jalm_dynamical_2019}, which identifies a breather solution to the NLS with a potential on a two-dimensional (2D) domain, but the mechanism associated with the breather remains unexplained. In general, very little is known about 2D breathers unless one considers discrete lattice problems.

In this letter, we demonstrate the existence of breather solutions to a family of (non-local) derivative NLS without a potential, realized in a 2D periodic domain (we call such solutions generally as breathers instead of quasi-breathers since the latter, as defined in \cite{pushkarev_quasibreathers_2013} for periodic domains, are associated with very different physics). In addition to being a novel 2D breather in a continuous field, other remarkable and distinguishing features of the solution include: (1) the breather spontaneously emerges from a stochastic wave field after long-time evolution, not relying on specific initial conditions; (2) the breather appears equivalently for both the focusing and defocusing cases, but exists only in the weak nonlinearity regime. As the nonlinearity of the system increases, we find a breakdown of the breather state with the field relaxing to the Rayleigh-Jeans spectrum (or wave turbulence spectrum if dissipation is included). Further analysis shows that the state trajectory of the breathers is associated with a Kolmogorov-Arnold-Moser (KAM) torus, which is a distorted trajectory of the linear (integrable) system that survives when the nonlinearity level is sufficiently low \cite{dumas_kam_2014}. 
%Such connections between KAM theory and breathers are not known in physics based on the authors' knowledge. 

\begin{figure*}
    \includegraphics{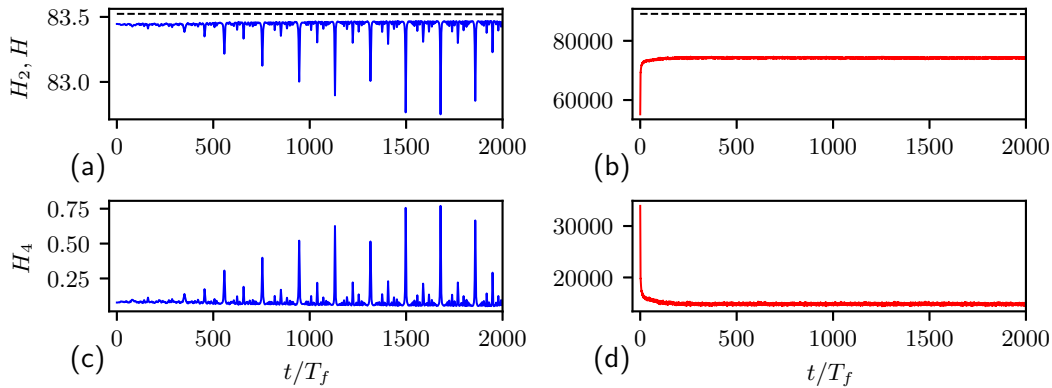}
    \caption{\label{fig:hamiltonian} The time series of $H$ (dashed) and $H_2$ (solid) for (a) $\varepsilon = 0.00071$ and (b) 
    $\varepsilon = 0.20$, as well as the corresponding $H_4$ for (c) $\varepsilon = 0.00071$ and (d) $\varepsilon = 0.20$. Note that a low sampling frequency (1 point per $T_f$) is used to plot the figure to improve its readability, leading to aliasing. Therefore, only a small portion of all $O(1600)$ peaks are visible. }
\end{figure*}

\section{Setup of Numerical Experiments}
The Majda-McLaughlin-Tabak (MMT) model is a family of nonlinear dispersive wave equations that have been widely used to study wave turbulence \cite{nazarenko_wave_2011,majda_one-dimensional_1997,zakharov_wave_2001} and coherent structures \cite{zakharov_wave_2001,rumpf_turbulent_2009,pushkarev_quasibreathers_2013,rumpf_wave_2013,rumpf_transition_2015}, due to its effectiveness in representing nonlinear waves in different physical contexts. In the present work, we consider the MMT model in two spatial dimensions, constructed as
\begin{equation}
i\frac{\partial \psi}{\partial t}=\vert\partial_{\bm{x}}\vert^{2}\psi+\lambda
\vert\partial_{\bm{x}}\vert^{\beta/4}\left(
\left|\vert\partial_{\bm{x}}\vert^{\beta/4}\psi\right|^{2}\vert\partial_{\bm{x}}\vert^{\beta/4}\psi \right),
\label{eqn:MMT}
\end{equation}
where $\psi\equiv\psi(\bm{x},t)$ is a complex scalar, $\bm{x}$ is the spatial coordinates, and $t$ the time. The non-local derivative operator $\vert\partial_{\bm{x}}\vert^\alpha$ denotes a multiplication by $k^\alpha$ on each spectral component in wave number domain, with $k=|\bm{k}|$. The free parameter $\beta$ controls the order of derivatives, and $\lambda=\pm 1$ generates a defocusing/focusing nonlinearity, respectively. Equation \eqref{eqn:MMT} is equivalent to a non-local derivative NLS, which can be shown more explicitly after a transformation $\phi=|\partial_{\bm{x}}|^{\beta/4}\psi$, leading to
\begin{equation}
i\frac{\partial \phi}{\partial t}=\vert\partial_{\bm{x}}\vert^{2}\phi+\lambda
\vert\partial_{\bm{x}}\vert^{\beta/2}\left(
\left|\phi\right|^{2}\phi \right).
\label{eqn:DNLS}
\end{equation}

The MMT model \eqref{eqn:MMT} can be derived from a Hamiltonian $H=H_{2}+H_{4}$, with
\begin{equation}
\begin{split}
H_2&=\int \big|\vert\partial_{\bm{x}}\vert\psi\big|^{2}d\bm{x}, \\
H_4&=\frac{1}{2}\lambda \int \left|\vert\partial_{\bm{x}}\vert^{\beta/4}\psi\right|^{4}d\bm{x}.
\label{eqn:Hamiltonian}
\end{split}
\end{equation}
The nonlinearity level of the system can be quantified via a parameter  $\varepsilon \equiv H_4 / H_2$.

We solve \eqref{eqn:MMT} on a 2D periodic domain, starting from an initial field $\psi_0\equiv\psi(\bm{x},t=0)$, via a pseudospectral method \cite{majda_one-dimensional_1997,hrabski_effect_2020,hrabski_energy_2021} with $128\times128$ modes (with higher resolution results available in supplemental material \cite{supp}). Our numerical method treats the linear term via an integrating factor, and the nonlinear term via an explicit 4th order Runge-Kutta scheme. The initial field $\psi_0$ is set as an exponential form in Fourier space as $\hat{\psi}_0 (\bm{k}) = A \text{exp}[-0.1|k-k_p|+i\theta_{\bm{k}}]$, where $k_p=4$, and $\theta_{\bm{k}}$ is a random phase that is decorrelated for all $\bm{k}$. In order to investigate dynamics at different nonlinearity levels, we choose a range of $A$ (about 10 values) leading to approximately $\varepsilon\in[0.0005,0.1]$ for each $\beta$ value of interest. 

\section{Results}
We start by describing a typical simulation leading to a breather state, with parameters $\beta = 3$ and $A=35$ (corresponding to $\varepsilon=0.00071$). Figures 1a and 1c show the long-time evolution of $H_2$ and $H_4$ from $t=0$ to $2000T_f$, with $T_f=2\pi$ the period of the fundamental wave mode. The total Hamiltonian $H$, as shown in figure 1a, is well conserved over $2000T_f$. After an initial evolution of about $400T_f$ with smooth profiles of $H_2$ and $H_4$, we observe that $H_4$ undergoes strong periodic jumps with corresponding dips in $H_2$. These jumps are associated with bursts of coherent structures, which are only present at low nonlinearity levels. In contrast, as demonstrated in figures 1b and 1d, the field evolution at a higher nonlinearity ($\varepsilon=0.20$) exhibits smooth profiles of $H_2$ and $H_4$ over the same time interval.

\begin{figure}[h]
    \includegraphics{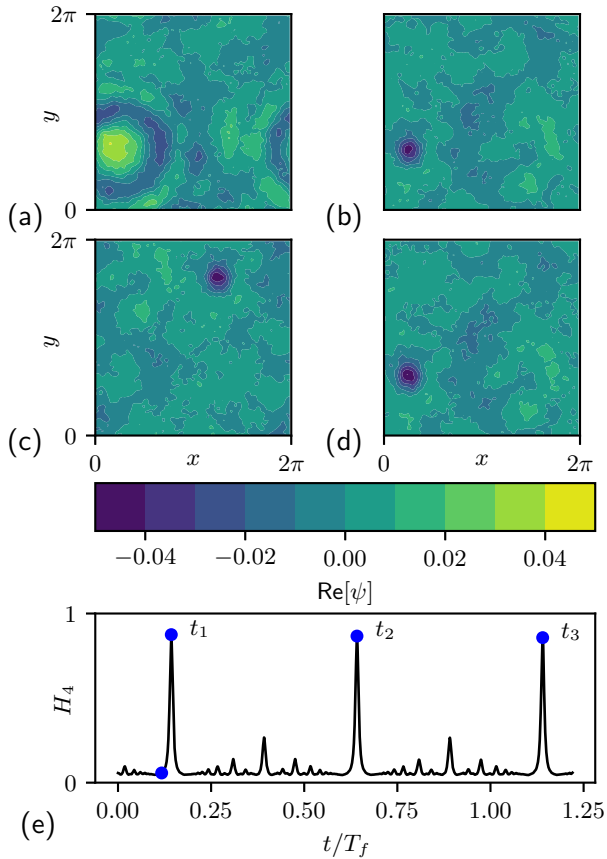}% Here is how to import EPS art
    \caption{\label{fig:psi} Contour plots of Re$[\psi]$ at $\beta=3$ for $\varepsilon=0.00071$ at various stages of the cycle of the breather (a/b/c/d), corresponding chronologically to the times marked by the blue circles in (e) the time series of $H_4$. Note that this plot of $H_4$ has sufficient sampling to resolve all features.}
\end{figure}

The oscillations in $H_4$ (and $H_2$) correspond to oscillations of a breather. To better visualize this breather state, we plot in figure \ref{fig:psi} the real part of $\psi$ at different phases of its oscillations (i.e., different stages of the oscillation pattern in $H_4$). Fig. \ref{fig:psi}a shows the field right before the first jump of $H_4$, where a concentric wave appears and later converges into a breather peak seen in Fig. \ref{fig:psi}b. This peak then collapses, with a second one emerging after about $T_f/2$ (according to fig. \ref{fig:psi}e) at the maximally distant location in the periodic domain, shown in \ref{fig:psi}c. The cycle then repeats itself with a peak emerging in Fig. \ref{fig:psi}d (at the same location as in \ref{fig:psi}a), forming an (oscillating) breather solution coexisting with a stochastic wave background. The smaller peaks of $H_4$ seen in Fig. \ref{fig:psi}e correspond to groups of secondary peaks in $\psi$. We encourage the reader to watch the animation of the full breather cycle (including the secondary peaks) in the supplemental material \cite{supp}. From figure \ref{fig:psi} we see that the breather oscillates with a period (very close to) $T_f$. Therefore, the simulation in figure 1a/c covers $O(1600)$ cycles of the breather, demonstrating a very long (perhaps infinite) time of existence. We note that only a small number of jumps are visible in figure 1a/c simply due to aliasing associated with limited resolution of plotting (see caption of figure \ref{fig:hamiltonian}).

\begin{figure}
    \includegraphics{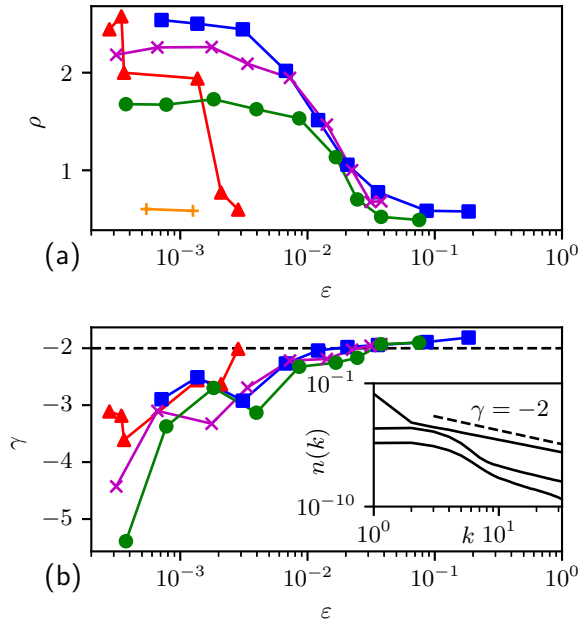}% Here is how to import EPS art
    \caption{\label{fig:gammarho} The quantities (a) $\rho$ and (b) $\gamma$ as functions of $\varepsilon$ for $\beta = 0$ (orange $+$), $\beta=1$ (green $\fullcirc$), $\beta=2$ (magenta $\times$), $\beta=3$ (blue  $\fullsquare$), and $\beta=4$ (red $\fulltriangle$). The inset of (b) shows fully-developed, angle-averaged wave action spectra at a few nonlinearity levels (for $\beta=3$), with the Rayleigh-Jeans spectral slope of $\gamma=-2$ indicated (dashed).}
\end{figure}

We next investigate the existence and intensity of the breathers for varying values of $\beta$ and $\epsilon$. To measure the intensity of the breather (relative to the background wave field), we define the peak-to-background ratio as
\begin{equation}
    \rho = \frac{\text{avgmax}[|\psi|]}{4\sigma_{|\psi|}},
    \label{eqn:rho}
\end{equation}
where the avgmax operator returns the average of the maximum height of primary peaks (as in Fig. \ref{fig:psi}b/c/d) over many cycles of the breather, and $\sigma_{|\psi|}$ is the total standard deviation of the field $|\psi|$ over space and time. By definition \eqref{eqn:rho}, $\rho=2$ corresponds to the typical rogue wave criterion used in many fields \cite{dudley_instabilities_2014}. 

Figure \ref{fig:gammarho}a shows the value of $\rho$ obtained for $\beta=0,1,2,3,4$ and $\varepsilon$ across three orders of magnitude. In general, we see that the breather state is present for smaller $\varepsilon$ (i.e., weak nonlinearity) and becomes stronger when $\beta$ is closer to 3. The case with $\beta=0$ (corresponding to NLS) leads to no breathers, indicating a non-local derivative in NLS is necessary for their emergence. We note that when the breathers are not present, the value of $\rho$ is evaluated by taking the average of the maximum of the field $|\psi|$ every $T_f/2$ as the numerator in \eqref{eqn:rho}.

Furthermore, we examine in figure 
\ref{fig:gammarho}b the slope $\gamma$ of the stationary wave action spectrum $n(\bm{k}) \equiv \langle \hat{\psi}(\bm{k}) \hat{\psi}^*(\bm{k}) \rangle$ across all values of $\beta$ and $\varepsilon$. The inset of figure \ref{fig:gammarho}b shows a typical example of fully-developed, angle-averaged $n(k)$ for $\beta=2$ and several values of $\varepsilon$. We see that the Rayleigh-Jeans spectrum with $\gamma=-2$ is only achieved at higher nonlinearity when the breather is not present. This trend is generally true for all values of $\beta$ as shown in Figure \ref{fig:gammarho}b.

A few additional remarks are in order. First, we note that the breather also emerges for the focusing equation \eqref{eqn:MMT} with $\lambda = -1$ under the same conditions. Second, the breather can also be observed under a forced/dissipated system \cite{hrabski_properties_2022}, but with the Rayleigh-Jeans spectrum replaced by the wave turbulence spectrum at high nonlinearity levels. Last but not least, we have performed extensive numerical analysis to verify that the breather we observe is not a numerical artifact. This includes the verification of the robustness of our results under symplectic integration, higher resolution, and different dealiasing schemes. Details of all of the above points can be found in the supplemental material \cite{supp}.

\section{Discussion}
In this section, we discuss the physical mechanism associated with the breather solution. We start by stating that there exists an exact breather solution to the linear system of \eqref{eqn:MMT}, i.e., $H_4=0$. What we mean precisely is that, starting from an initial condition with a breather peak (say figure \ref{fig:psi}b), the field propagated by the linear equation returns to the same state after exactly $T_f$. This is because the linear system only contains integer frequencies due to the NLS dispersion relation $\omega=k^2$, so that $T_f$ is the period of the linear system. This fact suggests that the breather solution to the nonlinear system arises from a deformed trajectory of the linear system.

\begin{figure}
    \includegraphics{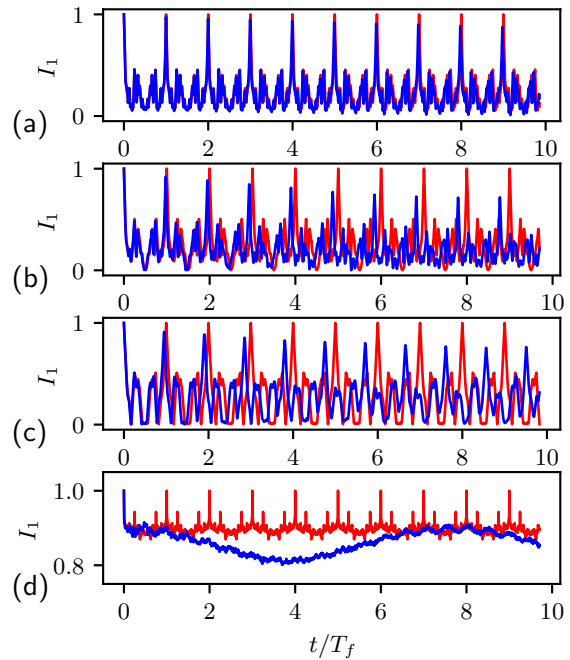}
    \caption{\label{fig:inner} $I_1$ evaluated for the nonlinear system $\beta=3$ (blue) and the linear system (red) for (a) $\varepsilon = 0.00071$, (b) $\varepsilon = 0.0084$, (c) $\varepsilon = 0.013$, and (d) $\varepsilon = 0.20$.}
\end{figure}

\begin{figure*}
    \includegraphics{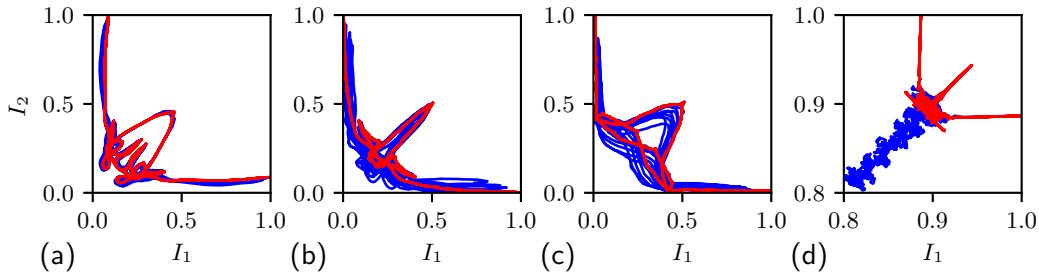}
    \caption{\label{fig:traj} The trajectory of $\psi$ projected on to $I_1$ and $I_2$ for several fundamental periods of $\psi$ computed via the nonlinear system $\beta=3$ (blue) and the linear system (red) for (a) $\varepsilon = 0.00071$, (b) $\varepsilon = 0.0084$, (c) $\varepsilon = 0.013$, and (d) $\varepsilon = 0.20$.}
\end{figure*}

Since visualizing the high-dimensional trajectory is very difficult, we define a projection of the trajectory to some physically meaningful reference field:
\begin{equation}
    I_n = \left|\frac{\int \psi^*_R(\bm{x},t_n ) \psi(\bm{x},t)  d\bm{x}}{\int \psi^*_R(\bm{x},t_n ) \psi_R (\bm{x},t_n)  d\bm{x}}\right|
    \label{eqn:inner}
\end{equation}
where $\psi^*_R(\bm{x},t_n)$ is the reference field where a breather peak is present, e.g., taken from $t_1$ in figure \ref{fig:psi}e, and $\psi(\bm{x},t)$ is the solution of either the linear or nonlinear system propagated from $\psi^*_R(\bm{x},t_n )$. Figure \ref{fig:inner} shows the evolution of $I_1$ from both linear and nonlinear systems for a range of four nonlinearity levels (at high nonlinearity in Fig. \ref{fig:inner}d, $t_1$ is taken from an arbitrary time in the quasi-stationary state). It is clear that the linear system evolution exhibits a period of $T_f$ in all sub-figures as expected. When the nonlinearity level is low, the trajectory identified by $I_1$ shows a small deformation from the linear trajectory, as seen in fig. \ref{fig:inner}a. Such deformation leads to a high-dimensional quasi-periodic trajectory that is evident from not only the gradual time shift of the $I_1$ peak, but also the deviation of $I_1$ peak from 1 (indicating completely different background waves). As nonlinearity level increases, we observe an increased deformation of the trajectory, until the linear trajectory is entirely destroyed at high nonlinearity in fig \ref{fig:inner}d. This trajectory deformation can also be observed in the $(I_1, I_2)$ plane in figure \ref{fig:traj} (on which the linear trajectory shows an ``L'' shape due to the orthogonality between $\psi^*_R(\bm{x},t_1)$ and $\psi^*_R(\bm{x},t_2)$), as well as in animations included the supplemental material \cite{supp}.

The trajectory deformation visualized above can be connected to Kolmogorov-Arnold-Moser (KAM) theory. If one considers the linear system as the base integrable system, then the nonlinear term is the perturbation added to the system to form a nearly integrable system. According to KAM theory, if the (nonlinear) perturbation is sufficiently small, some trajectories of linear system can be preserved with small deformation to a KAM torus \cite{dumas_kam_2014}. In our case, these preserved trajectories (i.e., KAM tori) correspond to those associated with the breather solution observed in figure \ref{fig:psi}. We also note that equation \eqref{eqn:MMT} (or more generally the NLS) is a special case on which KAM theory has not been fully understood mathematically, mainly because the base linear system does not contain irrational frequencies (or quasi-periodicity) that are required by the traditional KAM theorem. While most mathematical work of KAM on NLS relies on some way to introduce irrational frequencies, e.g., by including a potential term as in \cite{bourgain_quasi-periodic_1998}, recent progress \cite{geng_infinite_2011} does show that it is possible to prove the existence of quasi-periodic trajectories for the 2D NLS (without a potential) for small nonlinearity. Our results therefore demonstrate a breather solution associated with the quasi-periodic trajectory if a non-local derivative is included in the NLS. Furthermore, if a different dispersion relation, say $\omega=k^{1/2}$, is prescribed for \eqref{eqn:MMT}, the KAM theorem is more straightforward to apply (due to quasi-periodicity of the linear system). However, in this case, the breather state we observe is no longer a solution to the linear system, and one would not expect to observe it in the nonlinear system as well. We have confirmed this argument through numerical experiments.

Finally, the current analysis clearly does not resolve all the questions regarding the new 2D breather solution. One critical question is why it arises more strongly for $\beta$ close to 3. For these values of $\beta$, it is clear not only that the trajectories associated with breathers are stable, but also that they lie on some type of
quasi-periodic “attractor” \cite{lai_basins_2005}, such that a variety of initial conditions lead to the breather state. The reason for such dynamical behavior needs to be investigated in future work. In addition, our work provides a clear demonstration of the discrete wave turbulence regime governed by the KAM theory that has long been hypothesized in the literature \cite{lvov_discrete_2010}.

\section{Conclusion}
In this paper, we present results regarding a novel breather which spontaneously emerges from a 2D non-local derivative NLS. We show that the breather emerges at low nonlinearity with parameter $\beta$ close to 3. An analysis of phase space trajectory reveals that the trajectory associated with the breather solution is close to that of the linear system, but with quasi-periodicity introduced by the nonlinearity. The numerical findings support an explanation of the breather solution by the KAM theory, in the sense that a trajectory with the breather solution of the linear system is deformed but preserved when a small nonlinearity is introduced.  

\begin{acknowledgments}
We thank Benno Rumpf, Bobby Wilson, Peter Miller, Zaher Hani, Gigliola Staffilani, and Ricardo Grande for their thoughtful comments and suggestions. This material is based upon work supported by the National Science Foundation Graduate Research Fellowship under Grant No. DGE 1841052. Any opinions, findings, and conclusions or recommendations expressed in this material are those of the authors and do not necessarily reflect the views of the National Science Foundation. This work used the Extreme Science and Engineering Discovery Environment (XSEDE), which is supported by National Science Foundation grant number ACI-1548562. Computation was performed on XSEDE Bridges-2 at the Pittsburgh Supercomputing Center through allocation PHY200041. This research was supported in part through computational resources and services provided by Advanced Research Computing (ARC), a division of Information and Technology Services (ITS) at the University of Michigan, Ann Arbor.
\end{acknowledgments}

%\nocite{*}

\bibliography{HrabskiPan_Breather2022}% Produces the bibliography via BibTeX.

%apsrev4-2.bst 2019-01-14 (MD) hand-edited version of apsrev4-1.bst
%Control: key (0)
%Control: author (8) initials jnrlst
%Control: editor formatted (1) identically to author
%Control: production of article title (0) allowed
%Control: page (0) single
%Control: year (1) truncated
%Control: production of eprint (0) enabled
\begin{thebibliography}{32}%
\makeatletter
\providecommand \@ifxundefined [1]{%
 \@ifx{#1\undefined}
}%
\providecommand \@ifnum [1]{%
 \ifnum #1\expandafter \@firstoftwo
 \else \expandafter \@secondoftwo
 \fi
}%
\providecommand \@ifx [1]{%
 \ifx #1\expandafter \@firstoftwo
 \else \expandafter \@secondoftwo
 \fi
}%
\providecommand \natexlab [1]{#1}%
\providecommand \enquote  [1]{``#1''}%
\providecommand \bibnamefont  [1]{#1}%
\providecommand \bibfnamefont [1]{#1}%
\providecommand \citenamefont [1]{#1}%
\providecommand \href@noop [0]{\@secondoftwo}%
\providecommand \href [0]{\begingroup \@sanitize@url \@href}%
\providecommand \@href[1]{\@@startlink{#1}\@@href}%
\providecommand \@@href[1]{\endgroup#1\@@endlink}%
\providecommand \@sanitize@url [0]{\catcode `\\12\catcode `\$12\catcode
  `\&12\catcode `\#12\catcode `\^12\catcode `\_12\catcode `\%12\relax}%
\providecommand \@@startlink[1]{}%
\providecommand \@@endlink[0]{}%
\providecommand \url  [0]{\begingroup\@sanitize@url \@url }%
\providecommand \@url [1]{\endgroup\@href {#1}{\urlprefix }}%
\providecommand \urlprefix  [0]{URL }%
\providecommand \Eprint [0]{\href }%
\providecommand \doibase [0]{https://doi.org/}%
\providecommand \selectlanguage [0]{\@gobble}%
\providecommand \bibinfo  [0]{\@secondoftwo}%
\providecommand \bibfield  [0]{\@secondoftwo}%
\providecommand \translation [1]{[#1]}%
\providecommand \BibitemOpen [0]{}%
\providecommand \bibitemStop [0]{}%
\providecommand \bibitemNoStop [0]{.\EOS\space}%
\providecommand \EOS [0]{\spacefactor3000\relax}%
\providecommand \BibitemShut  [1]{\csname bibitem#1\endcsname}%
\let\auto@bib@innerbib\@empty
%</preamble>
\bibitem [{\citenamefont {Dysthe}\ and\ \citenamefont
  {Trulsen}(1999)}]{dysthe_note_1999}%
  \BibitemOpen
  \bibfield  {author} {\bibinfo {author} {\bibfnamefont {K.~B.}\ \bibnamefont
  {Dysthe}}\ and\ \bibinfo {author} {\bibfnamefont {K.}~\bibnamefont
  {Trulsen}},\ }\bibfield  {title} {\bibinfo {title} {Note on {Breather} {Type}
  {Solutions} of the {NLS} as {Models} for {Freak}-{Waves}},\ }\href
  {https://doi.org/10.1238/Physica.Topical.082a00048} {\bibfield  {journal}
  {\bibinfo  {journal} {Physica Scripta}\ }\textbf {\bibinfo {volume} {1999}},\
  \bibinfo {pages} {48} (\bibinfo {year} {1999})}\BibitemShut {NoStop}%
\bibitem [{\citenamefont {Onorato}\ \emph {et~al.}(2013)\citenamefont
  {Onorato}, \citenamefont {Proment}, \citenamefont {Clauss},\ and\
  \citenamefont {Klein}}]{onorato_rogue_2013}%
  \BibitemOpen
  \bibfield  {author} {\bibinfo {author} {\bibfnamefont {M.}~\bibnamefont
  {Onorato}}, \bibinfo {author} {\bibfnamefont {D.}~\bibnamefont {Proment}},
  \bibinfo {author} {\bibfnamefont {G.}~\bibnamefont {Clauss}},\ and\ \bibinfo
  {author} {\bibfnamefont {M.}~\bibnamefont {Klein}},\ }\bibfield  {title}
  {\bibinfo {title} {Rogue {Waves}: {From} {Nonlinear} {Schr{\"o}dinger}
  {Breather} {Solutions} to {Sea}-{Keeping} {Test}},\ }\href
  {https://doi.org/10.1371/journal.pone.0054629} {\bibfield  {journal}
  {\bibinfo  {journal} {PLOS ONE}\ }\textbf {\bibinfo {volume} {8}},\ \bibinfo
  {pages} {e54629} (\bibinfo {year} {2013})}\BibitemShut {NoStop}%
\bibitem [{\citenamefont {Dudley}\ \emph {et~al.}(2014)\citenamefont {Dudley},
  \citenamefont {Dias}, \citenamefont {Erkintalo},\ and\ \citenamefont
  {Genty}}]{dudley_instabilities_2014}%
  \BibitemOpen
  \bibfield  {author} {\bibinfo {author} {\bibfnamefont {J.~M.}\ \bibnamefont
  {Dudley}}, \bibinfo {author} {\bibfnamefont {F.}~\bibnamefont {Dias}},
  \bibinfo {author} {\bibfnamefont {M.}~\bibnamefont {Erkintalo}},\ and\
  \bibinfo {author} {\bibfnamefont {G.}~\bibnamefont {Genty}},\ }\bibfield
  {title} {\bibinfo {title} {Instabilities, breathers and rogue waves in
  optics},\ }\href {https://doi.org/10.1038/nphoton.2014.220} {\bibfield
  {journal} {\bibinfo  {journal} {Nature Photonics}\ }\textbf {\bibinfo
  {volume} {8}},\ \bibinfo {pages} {755} (\bibinfo {year} {2014})}\BibitemShut
  {NoStop}%
\bibitem [{\citenamefont {Ding}\ \emph {et~al.}(2019)\citenamefont {Ding},
  \citenamefont {Gao},\ and\ \citenamefont {Li}}]{ding_breathers_2019}%
  \BibitemOpen
  \bibfield  {author} {\bibinfo {author} {\bibfnamefont {C.-C.}\ \bibnamefont
  {Ding}}, \bibinfo {author} {\bibfnamefont {Y.-T.}\ \bibnamefont {Gao}},\ and\
  \bibinfo {author} {\bibfnamefont {L.-Q.}\ \bibnamefont {Li}},\ }\bibfield
  {title} {\bibinfo {title} {Breathers and rogue waves on the periodic
  background for the {Gerdjikov}-{Ivanov} equation for the {Alfv{\'e}n} waves
  in an astrophysical plasma},\ }\href
  {https://doi.org/10.1016/j.chaos.2019.01.007} {\bibfield  {journal} {\bibinfo
   {journal} {Chaos, Solitons \& Fractals}\ }\textbf {\bibinfo {volume}
  {120}},\ \bibinfo {pages} {259} (\bibinfo {year} {2019})}\BibitemShut
  {NoStop}%
\bibitem [{\citenamefont {Cuevas-Maraver}\ \emph {et~al.}(2014)\citenamefont
  {Cuevas-Maraver}, \citenamefont {Kevrekidis},\ and\ \citenamefont
  {Williams}}]{cuevas-maraver_sine-gordon_2014}%
  \BibitemOpen
  \bibinfo {editor} {\bibfnamefont {J.}~\bibnamefont {Cuevas-Maraver}},
  \bibinfo {editor} {\bibfnamefont {P.~G.}\ \bibnamefont {Kevrekidis}},\ and\
  \bibinfo {editor} {\bibfnamefont {F.}~\bibnamefont {Williams}},\ eds.,\ \href
  {https://doi.org/10.1007/978-3-319-06722-3} {\emph {\bibinfo {title} {The
  sine-Gordon Model and its Applications}}},\ \bibinfo {series} {Nonlinear
  Systems and Complexity}, Vol.~\bibinfo {volume} {10}\ (\bibinfo  {publisher}
  {Springer International Publishing},\ \bibinfo {address} {Cham},\ \bibinfo
  {year} {2014})\BibitemShut {NoStop}%
\bibitem [{\citenamefont {Tajiri}\ and\ \citenamefont
  {Watanabe}(1998)}]{tajiri_breather_1998}%
  \BibitemOpen
  \bibfield  {author} {\bibinfo {author} {\bibfnamefont {M.}~\bibnamefont
  {Tajiri}}\ and\ \bibinfo {author} {\bibfnamefont {Y.}~\bibnamefont
  {Watanabe}},\ }\bibfield  {title} {\bibinfo {title} {Breather solutions to
  the focusing nonlinear {Schr}{\textbackslash}"odinger equation},\ }\href
  {https://doi.org/10.1103/PhysRevE.57.3510} {\bibfield  {journal} {\bibinfo
  {journal} {Physical Review E}\ }\textbf {\bibinfo {volume} {57}},\ \bibinfo
  {pages} {3510} (\bibinfo {year} {1998})}\BibitemShut {NoStop}%
\bibitem [{\citenamefont {Clarke}\ \emph {et~al.}(2000)\citenamefont {Clarke},
  \citenamefont {Grimshaw}, \citenamefont {Miller}, \citenamefont
  {Pelinovsky},\ and\ \citenamefont {Talipova}}]{clarke_generation_2000}%
  \BibitemOpen
  \bibfield  {author} {\bibinfo {author} {\bibfnamefont {S.}~\bibnamefont
  {Clarke}}, \bibinfo {author} {\bibfnamefont {R.}~\bibnamefont {Grimshaw}},
  \bibinfo {author} {\bibfnamefont {P.}~\bibnamefont {Miller}}, \bibinfo
  {author} {\bibfnamefont {E.}~\bibnamefont {Pelinovsky}},\ and\ \bibinfo
  {author} {\bibfnamefont {T.}~\bibnamefont {Talipova}},\ }\bibfield  {title}
  {\bibinfo {title} {On the generation of solitons and breathers in the
  modified {Korteweg}{\textendash}de {Vries} equation},\ }\href
  {https://doi.org/10.1063/1.166505} {\bibfield  {journal} {\bibinfo  {journal}
  {Chaos: An Interdisciplinary Journal of Nonlinear Science}\ }\textbf
  {\bibinfo {volume} {10}},\ \bibinfo {pages} {383} (\bibinfo {year}
  {2000})}\BibitemShut {NoStop}%
\bibitem [{\citenamefont {Chow}\ \emph {et~al.}(2005)\citenamefont {Chow},
  \citenamefont {Grimshaw},\ and\ \citenamefont
  {Ding}}]{chow_interactions_2005}%
  \BibitemOpen
  \bibfield  {author} {\bibinfo {author} {\bibfnamefont {K.~W.}\ \bibnamefont
  {Chow}}, \bibinfo {author} {\bibfnamefont {R.~H.~J.}\ \bibnamefont
  {Grimshaw}},\ and\ \bibinfo {author} {\bibfnamefont {E.}~\bibnamefont
  {Ding}},\ }\bibfield  {title} {\bibinfo {title} {Interactions of breathers
  and solitons in the extended {Korteweg}{\textendash}de {Vries} equation},\
  }\href {https://doi.org/10.1016/j.wavemoti.2005.09.005} {\bibfield  {journal}
  {\bibinfo  {journal} {Wave Motion}\ }\textbf {\bibinfo {volume} {43}},\
  \bibinfo {pages} {158} (\bibinfo {year} {2005})}\BibitemShut {NoStop}%
\bibitem [{\citenamefont {Flach}\ and\ \citenamefont
  {Gorbach}(2008)}]{flach_discrete_2008}%
  \BibitemOpen
  \bibfield  {author} {\bibinfo {author} {\bibfnamefont {S.}~\bibnamefont
  {Flach}}\ and\ \bibinfo {author} {\bibfnamefont {A.~V.}\ \bibnamefont
  {Gorbach}},\ }\bibfield  {title} {\bibinfo {title} {Discrete breathers
  {\textemdash} {Advances} in theory and applications},\ }\href
  {https://doi.org/10.1016/j.physrep.2008.05.002} {\bibfield  {journal}
  {\bibinfo  {journal} {Physics Reports}\ }\textbf {\bibinfo {volume} {467}},\
  \bibinfo {pages} {1} (\bibinfo {year} {2008})}\BibitemShut {NoStop}%
\bibitem [{\citenamefont {Tr{\'i}as}\ \emph {et~al.}(2000)\citenamefont
  {Tr{\'i}as}, \citenamefont {Mazo},\ and\ \citenamefont
  {Orlando}}]{trias_discrete_2000}%
  \BibitemOpen
  \bibfield  {author} {\bibinfo {author} {\bibfnamefont {E.}~\bibnamefont
  {Tr{\'i}as}}, \bibinfo {author} {\bibfnamefont {J.~J.}\ \bibnamefont
  {Mazo}},\ and\ \bibinfo {author} {\bibfnamefont {T.~P.}\ \bibnamefont
  {Orlando}},\ }\bibfield  {title} {\bibinfo {title} {Discrete {Breathers} in
  {Nonlinear} {Lattices}: {Experimental} {Detection} in a {Josephson}
  {Array}},\ }\href {https://doi.org/10.1103/PhysRevLett.84.741} {\bibfield
  {journal} {\bibinfo  {journal} {Physical Review Letters}\ }\textbf {\bibinfo
  {volume} {84}},\ \bibinfo {pages} {741} (\bibinfo {year} {2000})}\BibitemShut
  {NoStop}%
\bibitem [{\citenamefont {Miroshnichenko}\ \emph {et~al.}(2001)\citenamefont
  {Miroshnichenko}, \citenamefont {Flach}, \citenamefont {Fistul},
  \citenamefont {Zolotaryuk},\ and\ \citenamefont
  {Page}}]{miroshnichenko_breathers_2001}%
  \BibitemOpen
  \bibfield  {author} {\bibinfo {author} {\bibfnamefont {A.~E.}\ \bibnamefont
  {Miroshnichenko}}, \bibinfo {author} {\bibfnamefont {S.}~\bibnamefont
  {Flach}}, \bibinfo {author} {\bibfnamefont {M.~V.}\ \bibnamefont {Fistul}},
  \bibinfo {author} {\bibfnamefont {Y.}~\bibnamefont {Zolotaryuk}},\ and\
  \bibinfo {author} {\bibfnamefont {J.~B.}\ \bibnamefont {Page}},\ }\bibfield
  {title} {\bibinfo {title} {Breathers in {Josephson} junction ladders:
  {Resonances} and electromagnetic wave spectroscopy},\ }\href
  {https://doi.org/10.1103/PhysRevE.64.066601} {\bibfield  {journal} {\bibinfo
  {journal} {Physical Review E}\ }\textbf {\bibinfo {volume} {64}},\ \bibinfo
  {pages} {066601} (\bibinfo {year} {2001})}\BibitemShut {NoStop}%
\bibitem [{\citenamefont {Livi}\ \emph {et~al.}(1997)\citenamefont {Livi},
  \citenamefont {Spicci},\ and\ \citenamefont {MacKay}}]{livi_breathers_1997}%
  \BibitemOpen
  \bibfield  {author} {\bibinfo {author} {\bibfnamefont {R.}~\bibnamefont
  {Livi}}, \bibinfo {author} {\bibfnamefont {M.}~\bibnamefont {Spicci}},\ and\
  \bibinfo {author} {\bibfnamefont {R.~S.}\ \bibnamefont {MacKay}},\ }\bibfield
   {title} {\bibinfo {title} {Breathers on a diatomic {FPU} chain},\ }\href
  {https://doi.org/10.1088/0951-7715/10/6/003} {\bibfield  {journal} {\bibinfo
  {journal} {Nonlinearity}\ }\textbf {\bibinfo {volume} {10}},\ \bibinfo
  {pages} {1421} (\bibinfo {year} {1997})}\BibitemShut {NoStop}%
\bibitem [{\citenamefont {Zakharov}\ \emph {et~al.}(2001)\citenamefont
  {Zakharov}, \citenamefont {Guyenne}, \citenamefont {Pushkarev},\ and\
  \citenamefont {Dias}}]{zakharov_wave_2001}%
  \BibitemOpen
  \bibfield  {author} {\bibinfo {author} {\bibfnamefont {V.~E.}\ \bibnamefont
  {Zakharov}}, \bibinfo {author} {\bibfnamefont {P.}~\bibnamefont {Guyenne}},
  \bibinfo {author} {\bibfnamefont {A.~N.}\ \bibnamefont {Pushkarev}},\ and\
  \bibinfo {author} {\bibfnamefont {F.}~\bibnamefont {Dias}},\ }\bibfield
  {title} {\bibinfo {title} {Wave turbulence in one-dimensional models},\
  }\href {https://doi.org/10.1016/S0167-2789(01)00194-4} {\bibfield  {journal}
  {\bibinfo  {journal} {Physica D: Nonlinear Phenomena}\ }\bibinfo {series}
  {Advances in {Nonlinear} {Mathematics} and {Science}: {A} {Special} {Issue}
  to {Honor} {Vladimir} {Zakharov}},\ \textbf {\bibinfo {volume} {152-153}},\
  \bibinfo {pages} {573} (\bibinfo {year} {2001})}\BibitemShut {NoStop}%
\bibitem [{\citenamefont {Rumpf}\ \emph {et~al.}(2009)\citenamefont {Rumpf},
  \citenamefont {Newell},\ and\ \citenamefont
  {Zakharov}}]{rumpf_turbulent_2009}%
  \BibitemOpen
  \bibfield  {author} {\bibinfo {author} {\bibfnamefont {B.}~\bibnamefont
  {Rumpf}}, \bibinfo {author} {\bibfnamefont {A.~C.}\ \bibnamefont {Newell}},\
  and\ \bibinfo {author} {\bibfnamefont {V.~E.}\ \bibnamefont {Zakharov}},\
  }\bibfield  {title} {\bibinfo {title} {Turbulent {Transfer} of {Energy} by
  {Radiating} {Pulses}},\ }\href
  {https://doi.org/10.1103/PhysRevLett.103.074502} {\bibfield  {journal}
  {\bibinfo  {journal} {Physical Review Letters}\ }\textbf {\bibinfo {volume}
  {103}},\ \bibinfo {pages} {074502} (\bibinfo {year} {2009})}\BibitemShut
  {NoStop}%
\bibitem [{\citenamefont {Pushkarev}\ and\ \citenamefont
  {Zakharov}(2013)}]{pushkarev_quasibreathers_2013}%
  \BibitemOpen
  \bibfield  {author} {\bibinfo {author} {\bibfnamefont {A.}~\bibnamefont
  {Pushkarev}}\ and\ \bibinfo {author} {\bibfnamefont {V.~E.}\ \bibnamefont
  {Zakharov}},\ }\bibfield  {title} {\bibinfo {title} {Quasibreathers in the
  {MMT} model},\ }\href {https://doi.org/10.1016/j.physd.2013.01.003}
  {\bibfield  {journal} {\bibinfo  {journal} {Physica D: Nonlinear Phenomena}\
  }\textbf {\bibinfo {volume} {248}},\ \bibinfo {pages} {55} (\bibinfo {year}
  {2013})}\BibitemShut {NoStop}%
\bibitem [{\citenamefont {Rumpf}\ and\ \citenamefont
  {Newell}(2013)}]{rumpf_wave_2013}%
  \BibitemOpen
  \bibfield  {author} {\bibinfo {author} {\bibfnamefont {B.}~\bibnamefont
  {Rumpf}}\ and\ \bibinfo {author} {\bibfnamefont {A.~C.}\ \bibnamefont
  {Newell}},\ }\bibfield  {title} {\bibinfo {title} {Wave instability under
  short-wave amplitude modulations},\ }\href
  {https://doi.org/10.1016/j.physleta.2013.03.015} {\bibfield  {journal}
  {\bibinfo  {journal} {Physics Letters A}\ }\textbf {\bibinfo {volume}
  {377}},\ \bibinfo {pages} {1260} (\bibinfo {year} {2013})}\BibitemShut
  {NoStop}%
\bibitem [{\citenamefont {Rumpf}\ and\ \citenamefont
  {Sheffield}(2015)}]{rumpf_transition_2015}%
  \BibitemOpen
  \bibfield  {author} {\bibinfo {author} {\bibfnamefont {B.}~\bibnamefont
  {Rumpf}}\ and\ \bibinfo {author} {\bibfnamefont {T.~Y.}\ \bibnamefont
  {Sheffield}},\ }\bibfield  {title} {\bibinfo {title} {Transition of weak wave
  turbulence to wave turbulence with intermittent collapses},\ }\href
  {https://doi.org/10.1103/PhysRevE.92.022927} {\bibfield  {journal} {\bibinfo
  {journal} {Physical Review E}\ }\textbf {\bibinfo {volume} {92}},\ \bibinfo
  {pages} {022927} (\bibinfo {year} {2015})}\BibitemShut {NoStop}%
\bibitem [{\citenamefont {Cretegny}\ \emph {et~al.}(1998)\citenamefont
  {Cretegny}, \citenamefont {Dauxois}, \citenamefont {Ruffo},\ and\
  \citenamefont {Torcini}}]{cretegny_localization_1998}%
  \BibitemOpen
  \bibfield  {author} {\bibinfo {author} {\bibfnamefont {T.}~\bibnamefont
  {Cretegny}}, \bibinfo {author} {\bibfnamefont {T.}~\bibnamefont {Dauxois}},
  \bibinfo {author} {\bibfnamefont {S.}~\bibnamefont {Ruffo}},\ and\ \bibinfo
  {author} {\bibfnamefont {A.}~\bibnamefont {Torcini}},\ }\bibfield  {title}
  {\bibinfo {title} {Localization and equipartition of energy in the
  $\beta$-{FPU} chain: {Chaotic} breathers},\ }\href
  {https://doi.org/10.1016/S0167-2789(98)00107-9} {\bibfield  {journal}
  {\bibinfo  {journal} {Physica D: Nonlinear Phenomena}\ }\textbf {\bibinfo
  {volume} {121}},\ \bibinfo {pages} {109} (\bibinfo {year}
  {1998})}\BibitemShut {NoStop}%
\bibitem [{\citenamefont {Rumpf}\ and\ \citenamefont
  {Newell}(2001)}]{rumpf_coherent_2001}%
  \BibitemOpen
  \bibfield  {author} {\bibinfo {author} {\bibfnamefont {B.}~\bibnamefont
  {Rumpf}}\ and\ \bibinfo {author} {\bibfnamefont {A.~C.}\ \bibnamefont
  {Newell}},\ }\bibfield  {title} {\bibinfo {title} {Coherent {Structures} and
  {Entropy} in {Constrained}, {Modulationally} {Unstable}, {Nonintegrable}
  {Systems}},\ }\href {https://doi.org/10.1103/PhysRevLett.87.054102}
  {\bibfield  {journal} {\bibinfo  {journal} {Physical Review Letters}\
  }\textbf {\bibinfo {volume} {87}},\ \bibinfo {pages} {054102} (\bibinfo
  {year} {2001})}\BibitemShut {NoStop}%
\bibitem [{\citenamefont {Rumpf}(2004)}]{rumpf_simple_2004}%
  \BibitemOpen
  \bibfield  {author} {\bibinfo {author} {\bibfnamefont {B.}~\bibnamefont
  {Rumpf}},\ }\bibfield  {title} {\bibinfo {title} {Simple statistical
  explanation for the localization of energy in nonlinear lattices with two
  conserved quantities},\ }\href {https://doi.org/10.1103/PhysRevE.69.016618}
  {\bibfield  {journal} {\bibinfo  {journal} {Physical Review E}\ }\textbf
  {\bibinfo {volume} {69}},\ \bibinfo {pages} {016618} (\bibinfo {year}
  {2004})}\BibitemShut {NoStop}%
\bibitem [{\citenamefont {Saint-Jalm}\ \emph {et~al.}(2019)\citenamefont
  {Saint-Jalm}, \citenamefont {Castilho}, \citenamefont {Le~Cerf},
  \citenamefont {Bakkali-Hassani}, \citenamefont {Ville}, \citenamefont
  {Nascimbene}, \citenamefont {Beugnon},\ and\ \citenamefont
  {Dalibard}}]{saint-jalm_dynamical_2019}%
  \BibitemOpen
  \bibfield  {author} {\bibinfo {author} {\bibfnamefont {R.}~\bibnamefont
  {Saint-Jalm}}, \bibinfo {author} {\bibfnamefont {P.~C.~M.}\ \bibnamefont
  {Castilho}}, \bibinfo {author} {\bibfnamefont {{\'E}.}~\bibnamefont
  {Le~Cerf}}, \bibinfo {author} {\bibfnamefont {B.}~\bibnamefont
  {Bakkali-Hassani}}, \bibinfo {author} {\bibfnamefont {J.-L.}\ \bibnamefont
  {Ville}}, \bibinfo {author} {\bibfnamefont {S.}~\bibnamefont {Nascimbene}},
  \bibinfo {author} {\bibfnamefont {J.}~\bibnamefont {Beugnon}},\ and\ \bibinfo
  {author} {\bibfnamefont {J.}~\bibnamefont {Dalibard}},\ }\bibfield  {title}
  {\bibinfo {title} {Dynamical {Symmetry} and {Breathers} in a
  {Two}-{Dimensional} {Bose} {Gas}},\ }\href
  {https://doi.org/10.1103/PhysRevX.9.021035} {\bibfield  {journal} {\bibinfo
  {journal} {Physical Review X}\ }\textbf {\bibinfo {volume} {9}},\ \bibinfo
  {pages} {021035} (\bibinfo {year} {2019})}\BibitemShut {NoStop}%
\bibitem [{\citenamefont {Dumas}(2014)}]{dumas_kam_2014}%
  \BibitemOpen
  \bibfield  {author} {\bibinfo {author} {\bibfnamefont {H.~S.}\ \bibnamefont
  {Dumas}},\ }\href@noop {} {\emph {\bibinfo {title} {The KAM Story: A Friendly
  Introduction to the Content, History, and Significance of Classical
  Kolmogorov{\textendash}Arnold{\textendash}Moser Theory}}}\ (\bibinfo
  {publisher} {WORLD SCIENTIFIC},\ \bibinfo {year} {2014})\BibitemShut
  {NoStop}%
\bibitem [{\citenamefont {Nazarenko}(2011)}]{nazarenko_wave_2011}%
  \BibitemOpen
  \bibfield  {author} {\bibinfo {author} {\bibfnamefont {S.}~\bibnamefont
  {Nazarenko}},\ }\href {https://doi.org/10.1007/978-3-642-15942-8} {\emph
  {\bibinfo {title} {Wave {Turbulence}}}},\ Lecture Notes in Physics\ (\bibinfo
   {publisher} {Springer-Verlag},\ \bibinfo {address} {Berlin Heidelberg},\
  \bibinfo {year} {2011})\BibitemShut {NoStop}%
\bibitem [{\citenamefont {Majda}\ \emph {et~al.}(1997)\citenamefont {Majda},
  \citenamefont {McLaughlin},\ and\ \citenamefont
  {Tabak}}]{majda_one-dimensional_1997}%
  \BibitemOpen
  \bibfield  {author} {\bibinfo {author} {\bibfnamefont {A.~J.}\ \bibnamefont
  {Majda}}, \bibinfo {author} {\bibfnamefont {D.~W.}\ \bibnamefont
  {McLaughlin}},\ and\ \bibinfo {author} {\bibfnamefont {E.~G.}\ \bibnamefont
  {Tabak}},\ }\bibfield  {title} {\bibinfo {title} {A one-dimensional model for
  dispersive wave turbulence},\ }\href {https://doi.org/10.1007/BF02679124}
  {\bibfield  {journal} {\bibinfo  {journal} {Journal of Nonlinear Science}\
  }\textbf {\bibinfo {volume} {7}},\ \bibinfo {pages} {9} (\bibinfo {year}
  {1997})}\BibitemShut {NoStop}%
\bibitem [{\citenamefont {Hrabski}\ and\ \citenamefont
  {Pan}(2020)}]{hrabski_effect_2020}%
  \BibitemOpen
  \bibfield  {author} {\bibinfo {author} {\bibfnamefont {A.}~\bibnamefont
  {Hrabski}}\ and\ \bibinfo {author} {\bibfnamefont {Y.}~\bibnamefont {Pan}},\
  }\bibfield  {title} {\bibinfo {title} {Effect of discrete resonant manifold
  structure on discrete wave turbulence},\ }\href
  {https://doi.org/10.1103/PhysRevE.102.041101} {\bibfield  {journal} {\bibinfo
   {journal} {Physical Review E}\ }\textbf {\bibinfo {volume} {102}},\ \bibinfo
  {pages} {041101} (\bibinfo {year} {2020})}\BibitemShut {NoStop}%
\bibitem [{\citenamefont {Hrabski}\ \emph {et~al.}(2021)\citenamefont
  {Hrabski}, \citenamefont {Pan}, \citenamefont {Staffilani},\ and\
  \citenamefont {Wilson}}]{hrabski_energy_2021}%
  \BibitemOpen
  \bibfield  {author} {\bibinfo {author} {\bibfnamefont {A.}~\bibnamefont
  {Hrabski}}, \bibinfo {author} {\bibfnamefont {Y.}~\bibnamefont {Pan}},
  \bibinfo {author} {\bibfnamefont {G.}~\bibnamefont {Staffilani}},\ and\
  \bibinfo {author} {\bibfnamefont {B.}~\bibnamefont {Wilson}},\ }\href
  {https://doi.org/10.48550/arXiv.2107.01459} {\bibinfo {title} {Energy
  transfer for solutions to the nonlinear {Schr}{\textbackslash}"odinger
  equation on irrational tori}} (\bibinfo {year} {2021})\BibitemShut {NoStop}%
\bibitem [{sup()}]{supp}%
  \BibitemOpen
  \href@noop {} {}\bibinfo {note} {See Supplemental Material at [URL] for
  demonstrations of the breather in additional contexts, plots of the secondary
  peaks of the breather, numerical verification of the results, and an
  animation of the breather.}\BibitemShut {Stop}%
\bibitem [{\citenamefont {Hrabski}\ and\ \citenamefont
  {Pan}(2022)}]{hrabski_properties_2022}%
  \BibitemOpen
  \bibfield  {author} {\bibinfo {author} {\bibfnamefont {A.}~\bibnamefont
  {Hrabski}}\ and\ \bibinfo {author} {\bibfnamefont {Y.}~\bibnamefont {Pan}},\
  }\bibfield  {title} {\bibinfo {title} {On the properties of energy flux in
  wave turbulence},\ }\bibfield  {journal} {\bibinfo  {journal} {Journal of
  Fluid Mechanics}\ }\textbf {\bibinfo {volume} {936}},\ \href
  {https://doi.org/10.1017/jfm.2022.106} {10.1017/jfm.2022.106} (\bibinfo
  {year} {2022})\BibitemShut {NoStop}%
\bibitem [{\citenamefont {Bourgain}(1998)}]{bourgain_quasi-periodic_1998}%
  \BibitemOpen
  \bibfield  {author} {\bibinfo {author} {\bibfnamefont {J.}~\bibnamefont
  {Bourgain}},\ }\bibfield  {title} {\bibinfo {title} {Quasi-{Periodic}
  {Solutions} of {Hamiltonian} {Perturbations} of {2D} {Linear}
  {Schr{\"o}dinger} {Equations}},\ }\href {https://doi.org/10.2307/121001}
  {\bibfield  {journal} {\bibinfo  {journal} {Annals of Mathematics}\ }\textbf
  {\bibinfo {volume} {148}},\ \bibinfo {pages} {363} (\bibinfo {year}
  {1998})}\BibitemShut {NoStop}%
\bibitem [{\citenamefont {Geng}\ \emph {et~al.}(2011)\citenamefont {Geng},
  \citenamefont {Xu},\ and\ \citenamefont {You}}]{geng_infinite_2011}%
  \BibitemOpen
  \bibfield  {author} {\bibinfo {author} {\bibfnamefont {J.}~\bibnamefont
  {Geng}}, \bibinfo {author} {\bibfnamefont {X.}~\bibnamefont {Xu}},\ and\
  \bibinfo {author} {\bibfnamefont {J.}~\bibnamefont {You}},\ }\bibfield
  {title} {\bibinfo {title} {An infinite dimensional {KAM} theorem and its
  application to the two dimensional cubic {Schr{\"o}dinger} equation},\ }\href
  {https://doi.org/10.1016/j.aim.2011.01.013} {\bibfield  {journal} {\bibinfo
  {journal} {Advances in Mathematics}\ }\textbf {\bibinfo {volume} {226}},\
  \bibinfo {pages} {5361} (\bibinfo {year} {2011})}\BibitemShut {NoStop}%
\bibitem [{\citenamefont {Lai}\ \emph {et~al.}(2005)\citenamefont {Lai},
  \citenamefont {He},\ and\ \citenamefont {Jiang}}]{lai_basins_2005}%
  \BibitemOpen
  \bibfield  {author} {\bibinfo {author} {\bibfnamefont {Y.-C.}\ \bibnamefont
  {Lai}}, \bibinfo {author} {\bibfnamefont {D.-R.}\ \bibnamefont {He}},\ and\
  \bibinfo {author} {\bibfnamefont {Y.-M.}\ \bibnamefont {Jiang}},\ }\bibfield
  {title} {\bibinfo {title} {Basins of attraction in piecewise smooth
  {Hamiltonian} systems},\ }\href {https://doi.org/10.1103/PhysRevE.72.025201}
  {\bibfield  {journal} {\bibinfo  {journal} {Physical Review E}\ }\textbf
  {\bibinfo {volume} {72}},\ \bibinfo {pages} {025201} (\bibinfo {year}
  {2005})}\BibitemShut {NoStop}%
\bibitem [{\citenamefont {L{\textquoteright}vov}\ and\ \citenamefont
  {Nazarenko}(2010)}]{lvov_discrete_2010}%
  \BibitemOpen
  \bibfield  {author} {\bibinfo {author} {\bibfnamefont {V.~S.}\ \bibnamefont
  {L{\textquoteright}vov}}\ and\ \bibinfo {author} {\bibfnamefont
  {S.}~\bibnamefont {Nazarenko}},\ }\bibfield  {title} {\bibinfo {title}
  {Discrete and mesoscopic regimes of finite-size wave turbulence},\ }\href
  {https://doi.org/10.1103/PhysRevE.82.056322} {\bibfield  {journal} {\bibinfo
  {journal} {Physical Review E}\ }\textbf {\bibinfo {volume} {82}},\ \bibinfo
  {pages} {056322} (\bibinfo {year} {2010})}\BibitemShut {NoStop}%
\end{thebibliography}%


%apsrev4-2.bst 2019-01-14 (MD) hand-edited version of apsrev4-1.bst
%Control: key (0)
%Control: author (8) initials jnrlst
%Control: editor formatted (1) identically to author
%Control: production of article title (0) allowed
%Control: page (0) single
%Control: year (1) truncated
%Control: production of eprint (0) enabled
\begin{thebibliography}{3}%
\makeatletter
\providecommand \@ifxundefined [1]{%
 \@ifx{#1\undefined}
}%
\providecommand \@ifnum [1]{%
 \ifnum #1\expandafter \@firstoftwo
 \else \expandafter \@secondoftwo
 \fi
}%
\providecommand \@ifx [1]{%
 \ifx #1\expandafter \@firstoftwo
 \else \expandafter \@secondoftwo
 \fi
}%
\providecommand \natexlab [1]{#1}%
\providecommand \enquote  [1]{``#1''}%
\providecommand \bibnamefont  [1]{#1}%
\providecommand \bibfnamefont [1]{#1}%
\providecommand \citenamefont [1]{#1}%
\providecommand \href@noop [0]{\@secondoftwo}%
\providecommand \href [0]{\begingroup \@sanitize@url \@href}%
\providecommand \@href[1]{\@@startlink{#1}\@@href}%
\providecommand \@@href[1]{\endgroup#1\@@endlink}%
\providecommand \@sanitize@url [0]{\catcode `\\12\catcode `\$12\catcode
  `\&12\catcode `\#12\catcode `\^12\catcode `\_12\catcode `\%12\relax}%
\providecommand \@@startlink[1]{}%
\providecommand \@@endlink[0]{}%
\providecommand \url  [0]{\begingroup\@sanitize@url \@url }%
\providecommand \@url [1]{\endgroup\@href {#1}{\urlprefix }}%
\providecommand \urlprefix  [0]{URL }%
\providecommand \Eprint [0]{\href }%
\providecommand \doibase [0]{https://doi.org/}%
\providecommand \selectlanguage [0]{\@gobble}%
\providecommand \bibinfo  [0]{\@secondoftwo}%
\providecommand \bibfield  [0]{\@secondoftwo}%
\providecommand \translation [1]{[#1]}%
\providecommand \BibitemOpen [0]{}%
\providecommand \bibitemStop [0]{}%
\providecommand \bibitemNoStop [0]{.\EOS\space}%
\providecommand \EOS [0]{\spacefactor3000\relax}%
\providecommand \BibitemShut  [1]{\csname bibitem#1\endcsname}%
\let\auto@bib@innerbib\@empty
%</preamble>
\bibitem [{\citenamefont {Rumpf}\ and\ \citenamefont
  {Newell}(2013)}]{rumpf_wave_2013}%
  \BibitemOpen
  \bibfield  {author} {\bibinfo {author} {\bibfnamefont {B.}~\bibnamefont
  {Rumpf}}\ and\ \bibinfo {author} {\bibfnamefont {A.~C.}\ \bibnamefont
  {Newell}},\ }\bibfield  {title} {\bibinfo {title} {Wave instability under
  short-wave amplitude modulations},\ }\href
  {https://doi.org/10.1016/j.physleta.2013.03.015} {\bibfield  {journal}
  {\bibinfo  {journal} {Physics Letters A}\ }\textbf {\bibinfo {volume}
  {377}},\ \bibinfo {pages} {1260} (\bibinfo {year} {2013})}\BibitemShut
  {NoStop}%
\bibitem [{\citenamefont {Hrabski}\ and\ \citenamefont
  {Pan}(2022)}]{hrabski_properties_2022}%
  \BibitemOpen
  \bibfield  {author} {\bibinfo {author} {\bibfnamefont {A.}~\bibnamefont
  {Hrabski}}\ and\ \bibinfo {author} {\bibfnamefont {Y.}~\bibnamefont {Pan}},\
  }\bibfield  {title} {\bibinfo {title} {On the properties of energy flux in
  wave turbulence},\ }\bibfield  {journal} {\bibinfo  {journal} {Journal of
  Fluid Mechanics}\ }\textbf {\bibinfo {volume} {936}},\ \href
  {https://doi.org/10.1017/jfm.2022.106} {10.1017/jfm.2022.106} (\bibinfo
  {year} {2022})\BibitemShut {NoStop}%
\bibitem [{\citenamefont {Sanz-Serna}\ and\ \citenamefont
  {Calvo}(2018)}]{sanz-serna_numerical_2018}%
  \BibitemOpen
  \bibfield  {author} {\bibinfo {author} {\bibfnamefont {J.~M.}\ \bibnamefont
  {Sanz-Serna}}\ and\ \bibinfo {author} {\bibfnamefont {M.~P.}\ \bibnamefont
  {Calvo}},\ }\href@noop {} {\emph {\bibinfo {title} {Numerical Hamiltonian
  Problems}}},\ \bibinfo {edition} {illustrated edition}\ ed.\ (\bibinfo
  {publisher} {Dover Publications},\ \bibinfo {address} {Mineola, New York},\
  \bibinfo {year} {2018})\BibitemShut {NoStop}%
\end{thebibliography}%

\end{document}

\title{Supplemental Material for the Paper \\ ``Breather Solutions to a Two-dimensional \\ Nonlinear Schr{\"o}dinger Equation with Non-local Derivatives''}

\author{Alexander Hrabski}
\author{Yulin Pan}
\affiliation{
Department of Naval Architecture and Marine Engineering, University of Michigan, Ann Arbor, Michigan 48109, USA
}

\maketitle
%\appendix
\section{Breather solutions in other situations}

%Additional Breather Examples}

In the main paper, we restrict our focus to a two-dimensional (2D), defocusing Majda-McLaughlin-Tabak (MMT) model without forcing or dissipation. In this section, we show the occurrence of the breather in additional contexts: the MMT model with a focusing nonlinearity, as well as a defocusing forced-dissipated model.

\begin{figure}
    \includegraphics{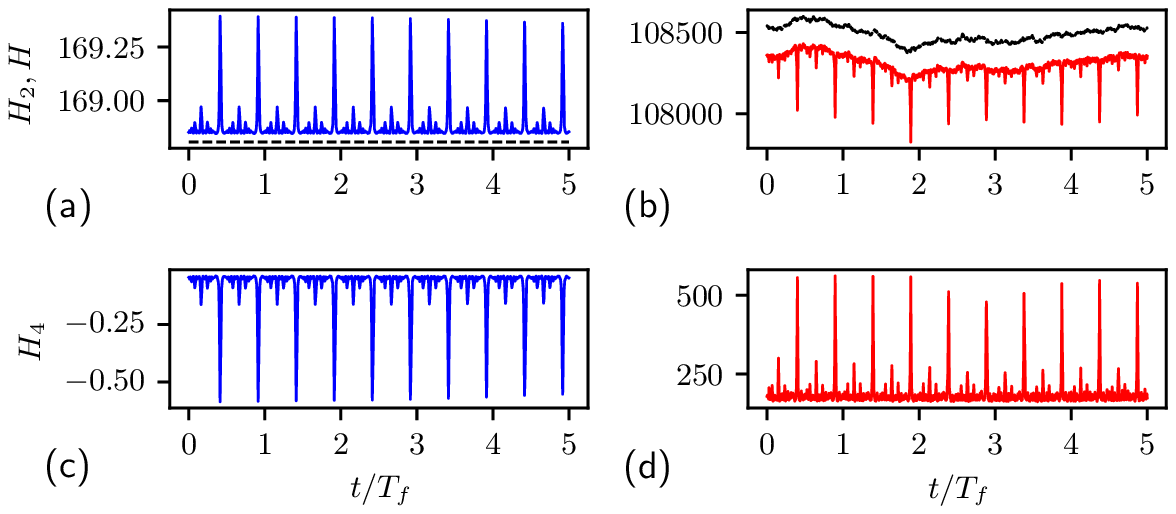}
    \caption{\label{fig:supp_hamiltonian} $5T_f$ of the time series of $H$ (dashed) and $H_2$ (solid) for a fully-developed breather solution to the (a) focusing MMT equation with $\beta=3$ and (b) forced-dissipated MMT equation with $\beta=2$. The corresponding $H_4$ for (c) the focusing system and (d) the forced-dissipated system are also provided. The focusing system has $\varepsilon = 0.00028$ and the forced-dissipated system has $\varepsilon = 0.0016$.}
\end{figure}

We begin with the focusing MMT model. The focusing case is given by equation (1) and (2) in the main paper with $\lambda = -1$. The parameter $\lambda$ is well-known to control the modulational instability of the Nonlinear Schr{\"o}dinger Equation (NLS) as well as the MMT model. In the context of the MMT model, the sign of $\lambda$ has been shown to affect the emergence of coherent structures in a one-dimensional MMT model with dispersion relation $\omega=k^{1/2}$ \cite{rumpf_wave_2013}. In our results, however, we find no significant change in the breather behavior between the focusing/defocusing equations, suggesting (along with the fact that the breather exists only at \emph{weak} nonlinearity) that modulational instability is not responsible for our breather. In figure \ref{fig:supp_hamiltonian}a/c, we show $5$ fundamental periods $T_f$ of $H$ and its components $H_2$ and $H_4$ in a fully-developed breather state for the focusing equation with $\beta = 2$. These results were obtained with an identical numerical setup to that of the main paper, and the presented results occur at low nonlinearity. The pattern of the oscillating breather is also similar to that in the defocusing case. 

\begin{figure}
    \includegraphics{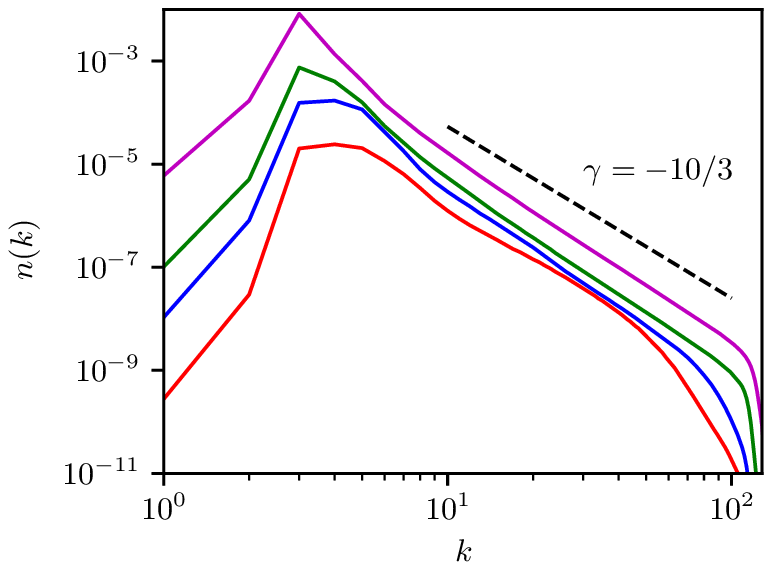}
    \caption{\label{fig:supp_spectra} The fully-developed, angle-averaged wave action spectra at a few nonlinearity levels of the forced-dissipated system, with the Kolmogorov-Zakharov spectral slope of $\gamma=-10/3$ indicated (dashed).}
\end{figure}

Next, we present results obtained for a forced-dissipated system. We again solve the defocusing 2D MMT model (with $\beta = 2$), however with the addition of forcing and dissipation terms. Specifically, we solve the equation
\begin{equation}
i\frac{\partial \psi}{\partial t}=\vert\partial_{\bm{x}}\vert^{2}\psi+\lambda
\vert\partial_{\bm{x}}\vert^{\beta/4}\left(
\left|\vert\partial_{\bm{x}}\vert^{\beta/4}\psi\right|^{2}\vert\partial_{\bm{x}}\vert^{\beta/4}\psi \right) + F + D_1 + D_2,
\label{eqn:supp_MMT_FD}
\end{equation}
where $F$ represents the forcing and $D_1$ and $D_2$ represent dissipation. These terms are explicitly defined in spectral domain, where 
\begin{equation}
    F = \left\{ \begin{array}{l}
    F_{r}+iF_{i} \ \ \ 7\le k\le9 \\
    0 \ \ \ \text{otherwise},
    \end{array} \right.
    \label{eqn:forcing}
\end{equation}
with $F_{r}$ and $F_{i}$ sampled from a Gaussian distribution of zero-mean, producing a standard white-noise forcing. The dissipative terms are defined as
\begin{eqnarray}
    D_1 =& \left\{ \begin{array}{l}
    -i\nu_1 \hat{\psi}_{\bm{k}} \ \ \ k\ge100 \\
    0 \ \ \ \text{otherwise},
    \end{array} \right. \nonumber \\
    D_2 =& \left\{ \begin{array}{l}
    -i\nu_2 \hat{\psi}_{\bm{k}} \ \ \ k\le7 \\
    0 \ \ \ \text{otherwise},
    \end{array} \right.
    \label{eqn:dissipation}
\end{eqnarray}
where $\nu_1$ and $\nu_2$ are dissipative constants. We solve these equations in an identical manner to that of the main paper with nearly identical initial conditions (we now begin with a spectral peak at $k_p = 10$) and on a larger domain of $512\times512$ modes. In figure \ref{fig:supp_hamiltonian}b, $H$ and its two components are plotted for $5T_f$ in the breather state. In this case, we do not expect the total Hamiltonian to be conserved, but rather to be quasi-steady for the fully developed system. Nevertheless, the signature of the breather is clearly present. Just as in the unforced case, we find that the wave action spectrum of the system is altered when the breather is present. In Figure \ref{fig:supp_spectra}, we provide the fully developed spectra of the forced-dissipated system for several different orders of nonlinearity. When the nonlinearity is low and the breather is present, we again see departure from a power-law spectrum, with a steeper tail region. When nonlinearity is raised, we observe the the spectra of wave turbulence are restored (and an associated forward energy cascade develops) \cite{hrabski_properties_2022}.

\section{secondary peaks in the breather cycle}
\begin{figure}
    \includegraphics{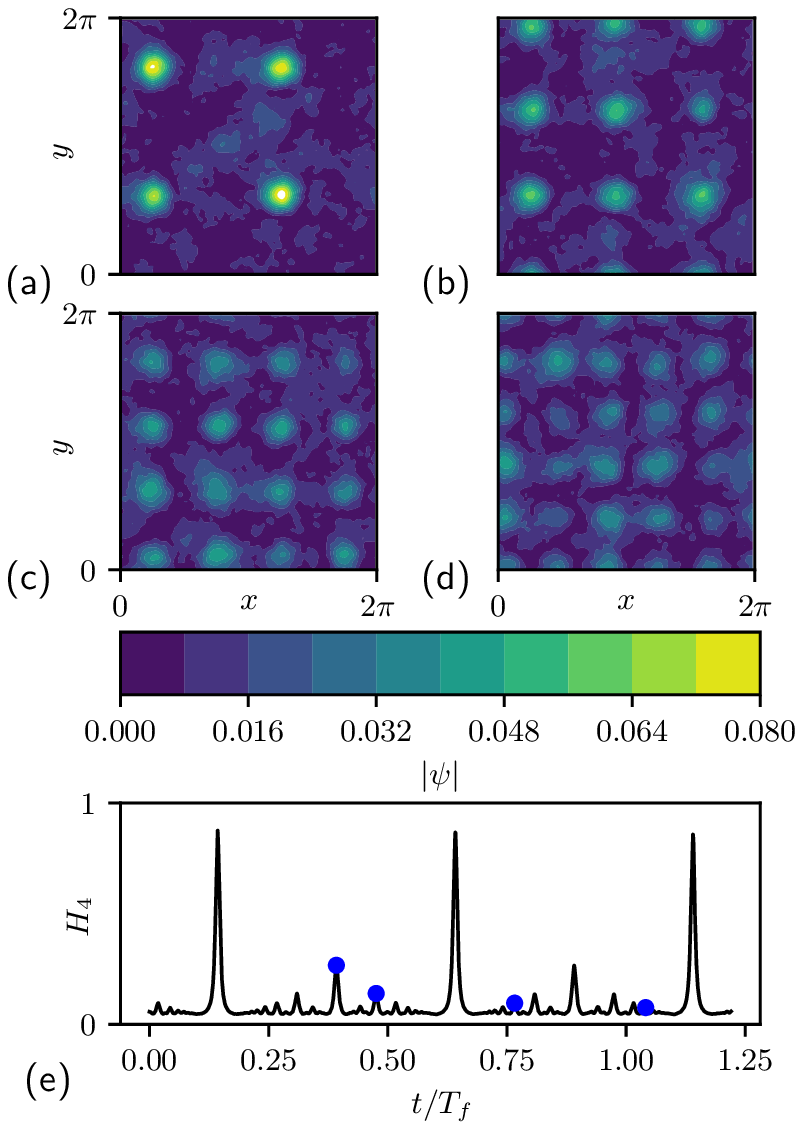}
    \caption{\label{fig:supp_psi} Contour plots of $|\psi|$ at $\beta=3$ for $\varepsilon=0.00071$ at various stages of the cycle of the breather (a/b/c/d), corresponding chronologically to the times marked by the blue circles in (e) the time series of $H_4$.}
\end{figure}

We include in this section plots of the secondary peaks of $|\psi|$ in the breather cycle, supplementing figure 2 in the main paper. We choose $|\psi|$ rather than Re$[\psi]$ (as in the main paper) to better resolve the smaller amplitudes of these secondary structures. While the largest peaks in $H_4$ (Fig. \ref{fig:supp_psi}e) correspond to single peaks in $|\psi|$, the secondary peaks in $H_4$ correspond to grids of smaller peaks in $|\psi|$ (Fig. \ref{fig:supp_psi}a/b/c/d).

\section{Numerical Validation of the Breather}

In this section, we provide analyses and numerical tests that rule out the possibility that the breather we discuss is a numerical artifact. In particular, we show that the breather solution is consistent under 
\begin{quote}
\begin{enumerate}[A.]
    \item the change of integration scheme to a symplectic integrator.
    \item the change of our dealiasing procedure.
    \item an increase in the number of Fourier modes (spatial resolution).
\end{enumerate}
\end{quote}

We begin with point A. Symplectic integration of a Hamiltonian system preserves the phase space geometry of its solution. Specifically, under Hamiltonian flow, structures such as sinks and limit cycles are forbidden by Liouville's theorem. When using an integrator such as an explicit 4th-order Runge-Kutta scheme (RK4), however, these structures can be erroneously introduced into the solution, which may change the dynamics. To ensure our breather is not an artifact introduced by non-symplectic integration, we implement a simple symplectic integrator, the implicit midpoint method (IMP) \cite{sanz-serna_numerical_2018}, to verify that we still obtain (and preserve) the breather solution. In the IMP method, we solve the implicit nonlinear problem via fixed-point iteration. For an identical numerical setup to that of the main paper, we allow the system to freely evolve under the IMP integration scheme. We set $\beta=3$ and simulate at the low nonlinearity of $\varepsilon = 0.001$.

\begin{figure}
    \includegraphics{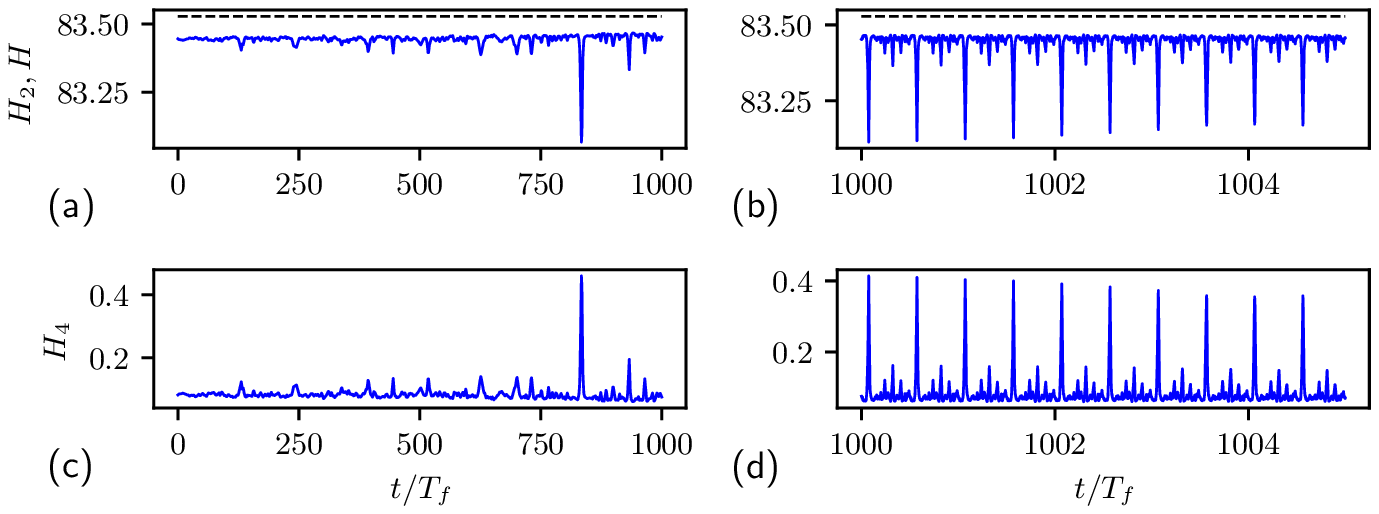}
    \caption{\label{fig:supp_symplectic} The time series of (a) $H$ (dashed) and $H_2$ (solid) starting from $t=0$ for the defocusing MMT equation with $\beta=3$ under symplectic integration, with the corresponding time series of (c) $H_4$. A detailed view beginning at $t=1000T_f$ of (b) $H$, $H_2$ and (d) $H_4$ over $5T_f$.}
\end{figure}

We provide in figure \ref{fig:supp_symplectic}a the evolution of $H$ and $H_2$ from $t=0$, with $H$ very well conserved and $H_2$ indicating that the breather has already formed by $t=1000T_f$. The corresponding plot of $H_4$ is provided in \ref{fig:supp_symplectic}c. Just as in the main paper, these plots of the initial evolution have a low sampling rate, leading to aliasing. To confirm that the breather has the same signature in $H_2$ and $H_4$ as in the case of non-symplectic integration, high-sampling rate plots of $H$,$H_2$ (Fig. \ref{fig:supp_symplectic}b) and $H_4$ (Fig. \ref{fig:supp_symplectic}d) are also provided over $5T_f$, showing no difference to those results of the main paper. Thus, the breather is not an artifact of non-symplectic integration.

Next, we address point B. In order to prevent the aliasing of modes due to the cubic nonlinearity of the MMT model, a standard $1/2$ dealiasing rule is applied after each product during the evaluation of the nonlinear term. The $1/2$ dealiasing rule is typically implemented via zero-padding the truncated wave number domain: if $k_m$ is the maximum resolved wave number  in our simulation (that is oriented along the $x$ and $y$ axes), then, in each direction, zero-padding is included such that for a computation domain of size $[-2k_m,2k_m]^2$, the non-zero (resolved) Fourier modes are only contained in the box $[-k_m,k_m]^2$. The zero-padding is enforced by setting all modes outside the box $[-k_m,k_m]^2$ to zero after each product is taken.

%the computation domain is of size $[-2k_m,2k_m]^2$, with the non-zero (resolved) Fourier modes contained in the box $[-k_m,k_m]^2$. The zero-padding is enforced by setting all modes with $|\bm{k}|_\infty > k_m$ to zero after each product is taken, where $|\bm{k}|_\infty = \max(k_x , k_y)$.

This procedure, however, has the effect of subtly changing the evolution of the system. In order to be assured that the breather is not an artifact of our dealiasing scheme, we first show that our dealiasing leads to a slightly modified Hamiltonian system (analytically), then we show that the breather is preserved in the original system without modification. We start by writing down the truncated Hamiltonian that we aim to numerically simulate:
\begin{equation}
H = \underset{\substack{\bm{k} \\ |\bm{k}|_\infty \in[-k_m ,k_m ]}}{\sum}k^2\hat{\psi}_{\bm{k}}\hat{\psi}_{\bm{k}}^* + \frac{1}{2}\lambda \underset{\substack{\bm{k}_1, \bm{k}_2, \bm{k}_3, \bm{k} \\ \bm{k}_1 + \bm{k}_2 = \bm{k}_3 + \bm{k} \\ |\bm{k}_i|_\infty \in[-k_m ,k_m ]}}{\sum}(k_1 k_2 k_3 k_4)^{\beta/4} \hat{\psi}_{\bm{k1}}\hat{\psi}_{\bm{k2}}\hat{\psi}_{\bm{k3}}^*\hat{\psi}_{\bm{k}}^*,
\label{eqn:supp_hamiltonian}
\end{equation}
where the summation is over every permutation over the subscript wave numbers. When computing the nonlinear term, we evaluate (via the Fourier transform)
\begin{equation}
    (\psi_{\bm{x}}\psi^*_{\bm{x}})\psi_{\bm{x}} = \left(\underset{\substack{\bm{k}_1, \bm{k}_3 \\ |\bm{k}_i|_\infty \in[-k_m ,k_m ] \\ {\color{red}|\bm{k}_1 - \bm{k}_3|_\infty \in[-k_m ,k_m ]}}}{\sum}\hat{\psi}_{\bm{k1}}\hat{\psi}^*_{\bm{k3}}e^{i(\bm{k1}-\bm{k3})\cdot\bm{x}} \right) \times \underset{\substack{\bm{k}_2 \\ |\bm{k}_2|_\infty \in[-k_m ,k_m ]}}{\sum}\hat{\psi}_{\bm{k2}}e^{i\bm{k2}\cdot\bm{x}}
    \label{eqn:supp_NLterm}
\end{equation}
where the derivatives have been neglected for clarity ($\beta=0$). The second condition under the first sum (red) is the first dealiasing step, where any product of modes that is mapped outside the bounded computational domain is excluded from the sum. The effect of dealiasing is therefore to remove certain interactions from the original system. It is not hard to show that including this extra condition modifies the Hamiltonian such that
\begin{equation}
H' = \underset{\substack{\bm{k} \\ |\bm{k}|_\infty \in[-k_m ,k_m ]}}{\sum}k^2\hat{\psi}_{\bm{k}}\hat{\psi}_{\bm{k}}^* + \frac{1}{2}\lambda \underset{\substack{\bm{k}_1, \bm{k}_2, \bm{k}_3, \bm{k} \\ \bm{k}_1 + \bm{k}_2 = \bm{k}_3 + \bm{k} \\ |\bm{k}_i|_\infty \in[-k_m ,k_m ] \\ {\color{red}|\bm{k}_1 - \bm{k}_3|_\infty \in[-k_m ,k_m ]}}}{\sum} (k_1 k_2 k_3 k_4)^{\beta/4} \hat{\psi}_{\bm{k1}}\hat{\psi}_{\bm{k2}}\hat{\psi}_{\bm{k3}}^*\hat{\psi}_{\bm{k}}^*,
\label{eqn:supp_hamiltonian_prime}
\end{equation}
where $H'$ represents the effective Hamiltonian when dealiasing is used. While a second dealiasing step is included after the second product is taken in \eqref{eqn:supp_NLterm}, no additional interactions are removed from $H'$ by the second dealiasing step: $|\bm{k}_1 + \bm{k}_2 - \bm{k}_3 |_\infty \in[-k_m ,k_m ]$ is accounted for by the fact that we already require $\bm{k}_1 + \bm{k}_2 - \bm{k}_3 = \bm{k}$ and $|\bm{k}|_\infty \in[-k_m ,k_m ]$. We remark that it is not $\emph{a priori}$ clear that the dealiased system is still Hamiltonian, but this fact is discovered when one attempts to write $H'$. 

In order to show that the system evolution according to \eqref{eqn:supp_hamiltonian} also leads to the breather solution, we perform a different dealiasing scheme for the simulation. Specifically, we skip the dealiasing step in the intermediate stage of computing the cubic term, and only dealias once after cubic multiplication is completed. Since this dealiasing step is equivalent to keeping only the Fourier modes up to $k_m$, this strategy produces evolution consistent with the system given by $H$ (rather than $H'$). We use this scheme in an otherwise identical setup to the main paper, with $\beta=3$ and $\varepsilon=0.001$, simulating until a breather emerges.

\begin{figure}
    \includegraphics{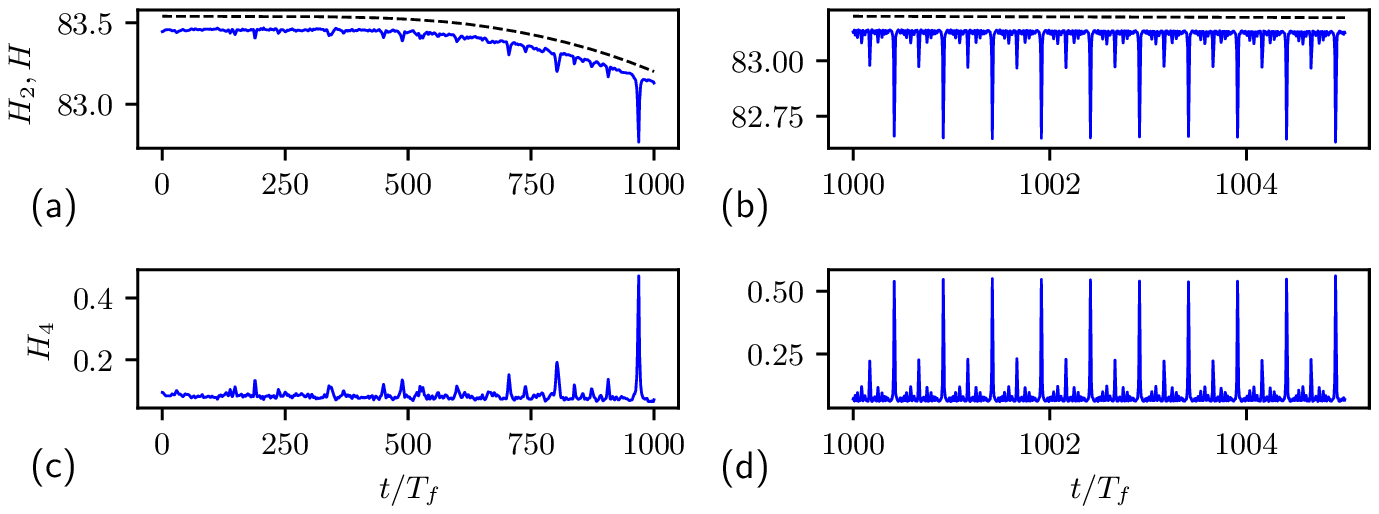}
    \caption{\label{fig:supp_dealias} The time series of (a) $H$ (dashed) and $H_2$ (solid) starting from $t=0$ for the defocusing MMT equation with $\beta=3$ using a scheme that avoids the dealiasing step, with the corresponding time series of (c) $H_4$. A detailed view beginning at $t=1000T_f$ of (b) $H$, $H_2$ and (d) $H_4$ over $5T_f$.}
\end{figure}

The evolution of the Hamiltonian $H$ and the component $H_2$ from $t=0$ are presented in figure \ref{fig:supp_dealias}a, and the corresponding $H_4$ in figure \ref{fig:supp_dealias}c. For this supplemental test we use a larger time step that leads to larger dissipation, though the energy loss over $1000T_f$ it is still only $0.5\%$ of the total energy. We see that a clear peak in $H_4$ has formed before $t=1000T_f$, indicating the breather has formed. Again, due to the low sampling rate, aliasing is present in the figures \ref{fig:supp_dealias}a and \ref{fig:supp_dealias}c. We provide detailed plots over $5T_f$ of $H$, $H_2$ in \ref{fig:supp_dealias}b and $H_4$ in \ref{fig:supp_dealias}d with sufficient sampling such that no aliasing is present. It is clear that the breather remains unchanged under our second scheme which preserves the truncated Hamiltonian system, indicating that the breather is not an artifact of dealiasing.

Finally, we address point C. The forced-dissipated results shown in \S I are computed on a domain with $16$ times as many modes, which shows that the breather emerges and persists in simulations with higher spatial resolution.

%of $4$ times the size of that of the main paper (with $16$ times as many modes), which shows that the breather emerges and persists 

%on larger domains in Fourier space (i.e., when spatial resolution is increased).

\bibliography{HrabskiPan_Breather2022}